%% file: 2014-mgmps-ARXIV.tex
\documentclass{article}
\usepackage{graphicx,url,a4wide}
\usepackage{amsmath,amssymb,amsfonts}
\input{00-symbols-ARXIV}

\begin{document}
\input{00-frontmatter-ARXIV}

\maketitle
\begin{abstract}
Computational theory and practice generally focus on single-paradigm systems, but relatively little is known about how best to combine components based on radically different approaches (e.g., silicon chips and wetware) into a single coherent system. In particular, while testing strategies for single-technology components are generally well developed, it is unclear at present how to perform integration testing on heterotic systems: can we develop a test-set generation strategy for checking whether specified behaviours emerge (and unwanted behaviours do not) when components based on radically different technologies are combined within a single system?

In this paper, we describe an approach to modelling multi-technology heterotic systems using a general-purpose formal specification strategy based on Eilenberg's $X$-machine model of computation. We show how this approach can be used to represent disparate technologies within a single framework, and propose a strategy for using these formal models for automatic heterotic test-set generation. We illustrate our approach by showing how to derive a test set for a heterotic system combining an $X$-machine-based device with a cell-based P system (membrane system).\\[12pt]

\end{abstract}

\input{01-sec-intro}

\input{02-sec-rev-testing}

\input{03-sec-rev-psystems}

\input{04-sec-example}

\input{05-sec-theory}
\input{06-sec-conclusions}

\bibliographystyle{alpha}
\bibliography{2014-mgmps-ptrsa}
\end{document}

%% file: 00-symbols-ARXIV.tex
\newtheorem{definition}{Definition}
\newtheorem{remark}{Remark}

\usepackage{xspace}

\newcommand{\MATH}[1]{\ensuremath{{#1}}\xspace}
\newcommand{\MATHBB}[1]{\MATH{\mathbb{#1}}}
\newcommand{\MATHCAL}[1]{\MATH{\mathcal{#1}}}
\newcommand{\MATHIT}[1]{\MATH{\mathit{#1}}}
\newcommand{\MATHSF}[1]{\MATH{\mathsf{#1}}}

\newcommand{\TUP}[1]{\MATHCAL{#1}}  
\newcommand{\OBJ}[1]{\MATHIT{#1}}   
\newcommand{\FUN}[1]{\MATHIT{#1}}   
\newcommand{\SET}[1]{\MATHSF{#1}}   

\newcommand{\Rset}{\MATHBB{R}}

\newcommand{\Inputs}{\SET{In}}
\newcommand{\Outputs}{\SET{Out}}
\newcommand{\States}{\SET{Q}}
\newcommand{\Mems}{\SET{Mem}}
\newcommand{\Procs}{\SET{Procs}}
\newcommand{\Starts}{\SET{Start}}
\newcommand{\Stops}{\SET{Stop}}
\newcommand{\InStreams}{\SET{\Inputs^{*}}}
\newcommand{\OutStreams}{\SET{\Outputs^{*}}}
\newcommand{\Tests}{\SET{Tests}}
\newcommand{\Cfgs}{\SET{Cfgs}}

\newcommand{\InPorts}{\SET{IN}}
\newcommand{\OutPorts}{\SET{OUT}}
\newcommand{\InPort}{\OBJ{in}}
\newcommand{\OutPort}{\OBJ{out}}

\newcommand{\Times}{\SET{Time}}

\newcommand{\Input}{\MATH{\iota}}
\newcommand{\Output}{\OBJ{o}}
\newcommand{\State}{\OBJ{q}}
\newcommand{\Mem}{\OBJ{m}}
\newcommand{\Proc}{\MATH{\varphi}}
\newcommand{\Start}{\OBJ{start}}
\newcommand{\Stop}{\OBJ{stop}}
\newcommand{\InStream}{\MATH{\sigma}}
\newcommand{\OutStream}{\MATH{\gamma}}
\newcommand{\NoSeq}{\MATH{\epsilon}} 

\newcommand{\NoMem}{\MATH{\lambda}} 
\newcommand{\Null}{\MATHSF{null}} 

\newcommand{\Ppoly}{\OBJ{P\!/\!poly}}
\newcommand{\Plogstar}{\OBJ{P\!/\!log^*}}

\newcommand{\Base}{\OBJ{Base}}
\newcommand{\Control}{\OBJ{Control}}

\newcommand{\mInit}{\MATH{\Mem\!^0}}

\newcommand{\Next}{\FUN{Next}}
\newcommand{\Trans}{\FUN{Trans}}

\newcommand{\XM}[1]{\TUP{#1}}  
\newcommand{\Aut}[1]{\TUP{#1}}  

\newcommand{\Spec}{\TUP{S}}
\newcommand{\Imp}{\TUP{I}}

\newcommand{\under}[1]{\FUN{\widehat{#1}}}

\newcommand{\where}{\MATH{\mathrel{~|~}}}
\newcommand{\Mapsto}{\MATH{\mathrel{~\longrightarrow~}}}
\newcommand{\Movesto}{\MATH{\mathrel{~\vdash~}}}

\newcommand{\Label}[3]{\MATH{{#2}/{#3}/{#1}}}
\newcommand{\Comp}[1]{\ensuremath{\mathop{ [\!|#1|\!]} }}

%% file: 00-frontmatter-ARXIV.tex
\title{Integration Testing of Heterotic Systems}
\author{Marian Gheorghe and Mike Stannett\\
Department of Computer Science\\
University of Sheffield\\
Regent Court, 211 Portobello, Sheffield S1 4DP\\
United Kingdom\\[6pt]
\url{{m.stannett,m.gheorghe}@sheffield.ac.uk}}
\date{11 August 2014}

%% file: 01-sec-intro.tex
\section{Introduction}
\label{sec:intro}

Modern technologies allow computation to be defined and implemented relative to a wide variety of paradigms and physical substrates, and it is natural to ask whether any advantage is to be gained by combining components based on radically different technologies to form a \emph{heterotic system}. Stepney et al. \cite{SAB+12} describe several instances of this idea, which at its most basic involves a system \XM{H} comprising two interacting components, \Base and \Control. The two components, possibly based on different computing paradigms, interact in a step-by-step manner. At each stage, the \Base component performs an action, thereby generating an output. This is interpreted by \Control, which then tells \Base what action to perform next.

The computational power of heterotic systems has been studied for many years. Towards the end of the twentieth century Siegelmann showed that no analogue device computing in polynomial time can compute more than the non-uniform complexity class \Ppoly \cite{Sie99}, while Bournez and Cosnard had previously argued that an idealised hybrid analogue/discrete dynamical system could in principle achieve this bound \cite{BC96}. More recent analyses by Tucker, Beggs and Costa have described a series of models that use experimental systems (\Base) as oracles providing data to an otherwise computable algorithm (\Control) -- the \Control layer observes the outcome of each \Base-level experiment, and uses this information to reconfigure \Base prior to the next experiment \cite{TB07}. Their results show that `interesting and plausible' model systems can, in principle, compute the smaller non-uniform complexity class \Plogstar, and they postulate \cite[p.~872]{BCT12} that this is essentially an upper limit for efficient real-world computation (``physical systems combined with algorithms cannot compute more in polynomial time than \Plogstar''). 

Kendon et al. \cite{KSS+11} have likewise pointed to the work of Anders and Browne \cite{AB09}, who observed that the combination of (efficiently) classically simulable \Control and \Base layers in a quantum cluster state computer results in a model which cannot be simulated efficiently. This implies that the interactions between two layers in a heterotic computer can contribute fundamentally to the power of the combined system, and this in turn has important consequences for anyone interested in the practicalities of testing such systems, since it tells us that the correctness of a heterotic system's behaviour cannot be assessed simply by examining the behaviours of its various components in isolation. While the components' correctness is obviously important, what Anders and Browne's example shows is that important aspects of a heterotic system's behaviour may depend not only on the components per se, but also on the intricate choreography of their interactions.

In this paper we focus on the complex question of integration testing, viz. how can we test the system obtained by combining \Base and \Control{}? We will illustrate our approach with a hybrid example drawn from the bio-related topic of P systems (membrane systems) \cite{handbook}.

\paragraph{Outline of paper.} In Sect. \ref{sec:rev:testing} we provide a review of $X$-machine testing strategies, which form the basis of our approach. In particular, we explain what a system of communicating stream X-machines (CSXMS) is, and how such a system can be tested. In Sect. \ref{sec:rev:psystems} we show how the CSXMS approach can be used to model and generate a test set for a heterotic system combining a stream X-machine (\Control) and a P system (\Base). To make this example accessible to readers, we first describe the biologically-based P system model in detail, and demonstrate how P system behaviours can themselves be unit tested. 

In Sect. \ref{sec:shortcomings} we identify shortcomings of our CSXMS testing approach, and discuss ongoing research into extending the underlying theory accordingly. We suggest in particular how a generalised theory of X-machine testing can be defined, which can be applied to heterotic systems in which the timing structures implicit in the system's behaviour are more complicated than allowed by existing approaches. Section \ref{sec:conclusion} concludes the paper, and includes suggestions for theoretical and experimental research towards validating the approach.

%% file: 02-sec-rev-testing.tex
\section{The X-machine testing methodology}
\label{sec:rev:testing}

In this section we introduce the basic concepts of the \emph{stream X-machine} (SXM) and \emph{communicating SXM} (CSXM), and describe what it means for an interacting collection of such machines to form a system (CSXMS). We explain what we mean by \emph{testing} such a system, and summarise the existing approach to SXM testing described in \cite{IH97,HI98}. Finally, we discuss a testing strategy for communicating SXM systems derived from the SXM testing methodology. For simplicity, we will only describe the procedures associated with testing deterministic machines, but a similar approach can also be developed for non-deterministic behaviours \cite{IH00}.

Stream X-machines were introduced by Laycock \cite{Lay93} as a variant of Eilenberg's \emph{$X$-machine} model of computation \cite{Eil74}, and we have recently described elsewhere how a generalised form of Eilenberg's original concept might be used to describe hybrid systems of unconventional computations \cite{Sta01,Sta14}. Our goal here is to expand on that description by showing in detail how the use of these models supports the identification of behavioural test-sets.

\paragraph{Notation.} Throughout this paper we write $\varnothing$ for the empty set and \Rset for the set of real numbers equipped with its standard algebraic and topological structures. Each natural number is interpreted to be the set of its predecessors, i.e. $0 \equiv \varnothing$, $n+1 \equiv \{ 0, 1, \dots, n \}$. In particular, we have $2 = \{ 0, 1 \}$.

If $X$ and $Y$ are sets, the set of total functions from $X$ to $Y$ is denoted $Y^X$. The domain of a function $f$ is denoted $dom(f)$. Since each subset $S$ of $X$ can be identified in terms of its characteristic function $\chi_S : X \to 2$, we write $2^X$ for the set of subsets of $X$ (the power set of $X$).

Given any set $X$, we define $X_\bot = X \cup \{ \bot \}$ where $\bot \not\in X$ is interpreted to mean `the undefined element of type $X$'. If ambiguity might otherwise arise, we write $\bot_X$ to indicate the set with which $\bot$ is associated. However, for historical reasons the `undefined memory' value (below) is generally called \NoMem instead of $\bot_\Mems$.

Given any alphabet $A$, we assume the existence of a symbol $\Null \not\in A$, with the property that prepending or appending \Null to any string in $A^*$ leaves that string unchanged, and likewise, if a variable $x$ is of type $A$, then the assignment $x \texttt{:=} \Null$ leaves the value of $x$ unchanged.

\subsection{Stream X-machines}
\label{sec:xm}

We recall the definition of a stream X-machine and some related concepts from \cite{HI98}.

\begin{definition}\label{def:xm}
A \emph{stream X-machine (SXM)} is a tuple
\[
   \XM{P} = (
	   \Inputs, \Outputs, \States, \Mems, 
		 \Procs, \Starts, \Stops, 
		 \mInit, \Next ) ,
\]
where
\begin{itemize}
\item 
	\Inputs and \Outputs are finite non-empty sets called the \emph{input alphabet} and \emph{output alphabet}, respectively, and \States is a finite non-empty set of \emph{states}; $\Starts \subseteq \States$ is the set of \emph{initial states} and $\Stops \subseteq \States$ is the set of \emph{terminal states};
\item 
	\Mems is a (possibly infinite) non-empty set of \emph{memory} values, and
	$\mInit \in \Mems$ is the \emph{initial memory};
\item 
	\Procs is a finite set of \emph{processing functions}. Each of these is of type $\Mems \times \Inputs \Mapsto \Outputs \times \Mems$;
\item 
	$\Next: \States \times \Procs \Mapsto 2^{\States}$ is a partial function, called the \emph{next-state function}.
\end{itemize}
\end{definition}

Intuitively, a stream $X$-machine can be regarded as a finite state machine \Aut{A}, equipped with transitions triggered by \Next and carrying labels of the form \Label{\Input}{\Output}{\Proc}, where $\Input \in \Inputs$, $\Output \in \Outputs$ and $\Proc \in \Procs$. Traversing such a transition is interpreted as consuming the input symbol \Input, updating the current memory from \Mem (say) to $\Proc(\Mem)$, and producing the output symbol \Output. We call \Aut{A} the automaton \emph{associated} with \XM{P}.  This process is \emph{deterministic} if \Starts contains just one element and \Next maps each state and processing function label onto at most one state, i.e. \Next can be regarded as a function $\Next: \States \times \Procs \Mapsto \States$. 
A \emph{configuration} of an SXM is a tuple $(\Mem,\State,\InStream,\OutStream)$, where 
  $\Mem \in \Mems$, $\State \in \States$, 
	$\InStream \in \InStreams$ and 
	$\OutStream \in \OutStreams$. It represents the idea that the machine is currently in state \State, the memory is currently \Mem, the machine's remaining input stream is \InStream, and it has so far produced the output stream \OutStream. An \emph{initial} configuration is one in which $\Mem=\mInit$, $\State \in \Starts$ and $\OutStream=\NoSeq$ (the empty sequence). A \emph{final} configuration has $\State \in \Stops$ and $\InStream=\NoSeq$.

We say that a \emph{configuration change}
$ 
   (\Mem,\State,\InStream,\OutStream) \Movesto 
	 (\Mem',\State',\InStream',\OutStream')
$
can occur provided
\begin{itemize}
\item $\InStream=\Input \InStream'$ for some $\Input \in \Inputs$;
\item $\OutStream'=\OutStream \Output$ for some $\Output \in \Outputs$; and 
\item there exists some $\Proc \in \Procs$ with $\State' \in \Next(\State,\Proc)$ and 
	$\Proc(\Mem,\Input) = (\Output, \Mem')$
\end{itemize}
The reflexive and transitive closure of \Movesto is denoted $\mathrel{\vdash\!^*}$. 

The \emph{relation computed} by an SXM \XM{M} is the relation $\Comp{M}: \InStreams \longleftrightarrow \OutStreams$ defined by
\begin{center} 
	$\InStream \Comp{M} \OutStream$ \\[6pt]
	iff there exist $\Start \in \Starts$, $\Stop \in \Stops$ and $\Mem\in \Mems$ such that\\[6pt]
	 $(\mInit, \Start, \InStream, \NoSeq) \mathrel{\vdash\!^*}
	  (\Mem, \Stop, \NoSeq, \OutStream)
	$.
\end{center}

\subsection{Communicating Stream X-machine Systems}
We introduce, loosely following \cite{BCGGHV99}, a simplified definition of communicating stream X-machines and communicating stream X-machine systems. A communicating SXM (CSXM) can be thought of as an SXM equipped with one input port (\InPorts) and one output port (\OutPorts). In a standard SXM, the next action of the machine in any given state (i.e. the processing function to be applied) is determined by the current input and current memory value. In a communicating SXM we also allow the machine to take into account the value, if any, currently present on the input port. The machine can also enter various special \emph{communicating} states, in which it transfers a memory value from its output port to the input port of another machine. This enables the various machines to exchange memory values as and when required, thereby allowing them to coordinate shared computations.

Notice that the input alphabet of a component machine $\Pi_i$ (the values which, together with its current memory, determine its behaviour) is a set of pairs, each describing the current input symbol and input port symbol (i.e. $\Inputs_i \times \InPorts_i$).\footnote{These definitions of $\Pi_i$'s input and outputs are technically only valid if each $\InPorts_i$ and $\OutPorts_i$ can be assumed finite. While we can rewrite the definition of a CSXM in a more rigorous, but more complicated, form to ensure that all input and output alphabets remain finite without regard to $\InPorts_i$ and $\OutPorts_i$, this is unnecessary for our purposes \cite{BCGGHV99}.} The result of firing a transition is more complex -- in addition to updating local memory the outcome can affect the local output stream, input port and output port, as well as the input port of any other machine in the system. Consequently, we take the output type to be $\Outputs_i \times \OutPorts_i \times \prod_{m=1}^n{\InPorts_m}$.

\begin{definition} 
\label{def:csxm}
\label{def:csxms}
A \emph{communicating stream X-machine system} (CSXMS) with $n$ components is an $n$-tuple
$
   \XM{P}_n = (\Pi_1, \dots, \Pi_n) ,
$
where each $\Pi_i$ is a \emph{communicating SXM} (CSXM), i.e. an SXM with input alphabet $\Inputs_i \times \InPorts_i$ and output alphabet $\Outputs_i \times \OutPorts_i \times \prod_{m=1}^n{\InPorts_m}$, where (writing $\Mems_i$ for the memory of $\Pi_i$, and similarly for its other components):
\begin{itemize}
\item 
  $\InPorts_i$ and $\OutPorts_i$ are both subsets of $({\Mems_i})_\bot$,
\item 
	$\States_i$ can be written as a disjoint union $\States_i=\States'_i\cup \States''_i$, where the elements of $\States'_i$ are called \emph{ordinary states} and those of $\States''_i$ are \emph{communicating states};
\item 
	$\Procs_i$ can be written as a disjoint union $\Procs_i=\Procs'_i\cup\Procs''_i$, where the elements of $\Procs'_i$ are called \emph{ordinary functions} and those of $\Procs''_i$ are \emph{communicating functions};
\item 
	The \emph{next-state function}, $\Next_i$, is undefined except on $(\States'_i\times \Procs'_i) \cup (\States''_i\times \Procs''_i)$, and $\Next_i(\State'',\Proc'')\subseteq \States'_i$ for all $\State''\in \States''_i$, $\Proc''\in\Procs''_i$, i.e., ordinary states support ordinary functions, communicating states support communicating function, and the target state of a communicating function is always an ordinary state.
\end{itemize}
\end{definition}

\paragraph{Configurations and configuration changes in a CSXMS.}
A \emph{configuration} of a component CSXM $\Pi_i$ is a tuple $c_i = (\Mem, \State, \InStream, \OutStream,\InPort,\OutPort)$, where $\InPort \in \InPorts_i$, $\OutPort \in \OutPorts_i$, and the other entries are defined as before. Given a CSXMS $\XM{P}_n = (\Pi_1, \dots, \Pi_n)$, we define a \emph{configuration} of $\XM{P}_n$ to be a tuple $(c_1, \dots, c_n)$ where each $c_i$ is a configuration of the corresponding $\Pi_i$. A configuration $(c_1, \dots, c_n)$ is \emph{initial} provided each $c_i$ is initial (including the requirement that $\InPort = \OutPort = \NoMem$, so that the first move made by the machine must be ordinary).

There are two ways in which a CSXMS can change its configuration. An \emph{ordinary} configuration change is one that causes no communication between machines; each machine can either consume and process a symbol present on the input channel, or it can leave the input channel untouched. We model this second case by saying that it consumes the undefined \NoMem symbol. A \emph{communicating} configuration change is one in which a symbol is removed from one machine's output port and a corresponding symbol is inserted into a second machine's output port, provided it is currently empty.

\begin{definition}\label{csxms-config-change}
A configuration change $(c_1, \dots, c_n) \vdash (c'_1, \dots, c'_n)$ is \emph{ordinary} if there is some $i$ such that $c'_j = c_j$ for all $j \neq i$, and some ordinary function $\Proc' \in \Procs'_i$ with
\begin{itemize}
\item $\State' \in \Next_i(\State,\Input,\InPort)$,
\item $\OutPort' \in \OutPorts_i$, and either
\begin{itemize}
\item $\Proc'(\Input,\InPort,\Mem) = (\Mem', \Output',\OutPort',\langle{\InPort_m'}\rangle_{m=1}^n)$, where $\InPort \neq \NoMem$ and $\InPort' = \NoMem$;
\item $\Proc'(\Input,\NoMem,\Mem)  = (\Mem', \Output',\OutPort',\langle{\InPort_m'}\rangle_{m=1}^n)$ and $\InPort' = \InPort$.
\end{itemize}
\end{itemize}
It is \emph{communicating} if there exists some $i \neq k$ such that 
\begin{itemize}
\item $c'_j = c_j$ for all $j \not\in \{i,k\}$,
\item $\OutPort_k = \OutPort'_i = \NoMem$,
\item $\State'_i \in \States'_i$ (the next state in the sending machine is ordinary),
\item $\InStream'_i = \InStream_i$, $\OutStream'_i = \OutStream_i$, $\InStream'_k = \InStream_k$, $\OutStream'_k = \OutStream_k$ (all input and output streams are unchanged)
\item there exists some communicating function $\Proc'' \in \Procs''_i$ which can be applied in the current state, and which generates the symbol that appears in the target machine's input port, i.e.
\begin{itemize}
\item $\State'_i \in \Next_i(\State_i,\Input_i,\InPort_i)$, and
\item $\Proc'(\Input_i,\InPort_i,\Mem_i) = (\Mem'_i, \Null,\NoMem,\langle{\InPort_m'}\rangle_{m=1}^n)$
\end{itemize}
\end{itemize}
where $c_i =  (\Mem_i, \State_i, \InStream_i, \OutStream_i,\InPort_i,\OutPort_i)$, etc.
\end{definition}


\begin{remark}\label{rem:csxms}
A CSXMS, $\XM{P}_n = (\Pi_1, \dots, \Pi_n)$, functions as follows:
\begin{itemize}
\item[(i)] each $\Pi_i$ starts with both the input and output ports containing $\NoMem$. The only function that can be applied initially should be an ordinary processing function, $\Proc_i \in \Procs'_i$. Hence, the initial state $\State^0\in \Starts_i$ from which $\Proc_i$ emerges must be an ordinary state;
\item[(ii)] an ordinary function $\Proc_i$ can process a symbol from $\InPorts_i$ if one is present, or it can proceed by ignoring the input value in which case the content of $\InPorts_i$ remains unchanged. A similar behaviour is expected for the $\OutPorts_i$ port;
\item[(iii)] after a communicating function is applied, the machine state will be an ordinary one, and so the next function to be applied (if any) will also be ordinary.
\end{itemize}
\end{remark}

\subsection{X-machine Testing}
\label{sec:rev:xmtesting}

The fact that an SXM can be regarded as an augmented version of its associated automaton means that well established automated finite state machine test-set generation strategies (e.g., based on Chow's W-method \cite{Cho78}) can be `lifted' to provide SXM testing strategies. The goal of SXM testing is to establish whether two SXMs, \Spec (the \emph{specification}) and \Imp (the \emph{implementation under test}, or \emph{IUT}) compute the same behaviour. We assume that the complete structure of \Spec is known and that \Spec has been minimised, that \Spec and \Imp use the same set \Procs of processing functions (if not, we define \Procs to be the union of their respective process sets), and attempt to find a finite \emph{test set}, $\Tests \subset \InStreams$, with the property that, if $\Comp{\Spec}(t) = \Comp{\Imp}(t)$ for every $t \in \Tests$, then \Spec and \Imp must necessarily compute the same relation. In general, the ability to store data in memory during a computation means that this problem is well-known to be uncomputable; it is therefore necessary to impose certain constraints, called \emph{design for test} (DFT) conditions, as to which implementations \Imp are considered valid candidates for testing. In particular, we generally assume that some estimate is available as to how many extra states \Imp has relative to \Spec.

DFT conditions for stream X-machines are well known, and an adequate set of conditions to ensure testability is \cite{HI98}:
\begin{itemize}
\item 
	\textbf{deterministic specification}: the behaviours of \Imp and its associated automaton $A$ should both be deterministic, i.e., given any state and any two processing functions, $\Proc_1$ and $\Proc_2$, applicable in that state, we require $dom(\Proc_1) \cap dom(\Proc_2) = \varnothing$;
\item 
	\textbf{$\Procs$-completeness}: 
	given any $\Proc\in \Procs$ and $\Mem\in \Mems$, there exists some $\Input\in \Inputs$ such that $\Proc(\Mem,\Input)$ is defined;
\item 
	\textbf{$\Procs$-output distinguishability}: examining the output of a processing function should tell us which function it is, i.e. given any $\Proc_1, \Proc_2 \in \Procs$, if there exist $\Mem, \Mem_1, \Mem_2 \in \Mems, \Input \in \Inputs$ and $\Output \in \Outputs$ such that $\Proc_1(\Mem,\Input)=(\Output,\Mem_1)$ and $\Proc_2(\Mem,\Input)=(\Output,\Mem_2)$, then $\Proc_1=\Proc_2$.
\end{itemize}

Since the SXM testing methodology requires us to examine the outputs that are produced when certain test inputs are processed, extending the technique to include CSXM systems requires the designer to ensure that every function application consumes an input and produces an output. As the communicating functions act only on memory symbols these must therefore be extended to handle input and output symbols. To do this, an additional input symbol $a \notin \bigcup{\Inputs_i}$ is introduced and for each communicating function $\Proc''_j \in \Procs_i''$ an output symbol $[i,j]$ is added. We now formally redefine $\Proc''_j$ to take the input symbol $a$ (\textit{this is a communication event}) and generate the output symbol $[i,j]$ (\textit{I have just applied $\Pi_i$'s communication function, $\Proc''_j$}). As before, each component CSXM, $\Pi_i$, should be deterministic, $\Procs_i$-complete and $\Procs_i$-output distinguishable (the extensions applied to the communicating functions mean that these automatically satisfy the last two conditions). The entire CSXMS, $\XM{P}_n$, is then converted into a single SXM, \XM{P^T}, and standard SXM testing is applied; however, although the CSXM components are deterministic, the resulting SXM need not be and consequently a testing approach for non-deterministic SXMs is used \cite{IH00}. 

The SXM, $\XM{P^T}=(\Inputs,\Outputs,\States,\Mems,\Procs,\Next,\Starts,\Stops,\mInit)$, is obtained from the CSXMS, $\mathcal{P}_n$, with the additional extensions mentioned above, as follows \cite{IH02}:\footnote{A similar testing approach is proposed in \cite{IBE03} for a slightly different CSXMS concept.} 
\begin{itemize}
\item 
  $\Inputs = 
	  \left( (\Inputs_1\cup \{a,\Null\}) \times \dots \times (\Inputs_n\cup \{a,\Null\}) \right)
		\setminus \{(\Null, \dots, \Null)\}$
\item 
  $\Outputs = ((\Outputs_1\cup \{[1,j]| j \neq 1\} \cup\{\Null\})$ 
	   $\times \dots$ \\ 
		\text{\qquad\qquad} $\dots$ $\times$ 
		          $(\Outputs_n\cup \{[n,j]| j \neq n\} \cup\{\Null\}))$ 
		$\setminus$ $\{(\Null, \dots, \Null)\}$
\item 
  $\States=\States_1\times\dots \times \States_n$, \quad
	$\Starts=I_1\times\dots \times I_n$, \quad
	$\Stops=T_1 \times\dots\times T_n$
\item 
	$\Mem=(\InPorts_1\times \Mems \times \OutPorts_1) \times \dots \times (\InPorts_n\times \Mems \times \OutPorts_n)$.
\item 
	$\mInit=((\NoMem, \mInit_1,\NoMem), \dots, (\NoMem, \mInit_n,\NoMem))$.
\item 
  $\Procs =\{ (\overline{\Proc_1}, \dots, \overline{\Proc_n}) \} \where (\forall i)(\overline{\Proc_i} \in \Procs_i \cup \{ \mathrm{id}_i \}) \}$
\end{itemize}
The SXM $\XM{P^T}$ is the product of the CSXMS components. A processing function, $\Proc$, describes a set of functions that are simultaneously applied in the CSXMS components. However, some components might not execute any processing functions during a particular computation step; in this case $\overline{\Proc_i}=e$.

The associated test set consists of input sequences obtained by applying a so-called \emph{fundamental test function}, $t: \Procs^* \Mapsto \InStreams$, to a sequence of processing functions derived from the associated automaton by applying one of the many known state machine based testing methods \cite{LY96}. Formally, a \emph{test set} for an SXM is a finite set of input sequences 
\[
	\Tests =\{ \Input_1\dots \Input_p \in \InStreams
	  \where
		\exists \Proc_1\dots \Proc_p \in \Procs \quad\text{ s.t. }\quad 
		        t(\Proc_1 \dots \Proc_p) = \Input_1 \dots \Input_p \},
\]
where we require, for each processing function $f_i=(\overline{\Proc_{i,1}}, \dots, \overline{\Proc_{i,n}})$ and each associated input element, $\Input_i = (\overline{\Input_{i,1}}, \dots, \overline{\Input_{i,n}})$, that
\begin{itemize}
\item
  when $\overline{\Proc_{i,j}}$ is either an ordinary or communicating function, then $\overline{\Input_{i,j}} \in (\Inputs_j \cup \{a\})$; 
\item
	otherwise, when $\overline{\Proc_{i,j}}= \mathrm{id}_j$ then $\overline{\Input_{i,j}}=\null$ (i.e., when the current configuration of the $j-$th component remains unchanged, then there is no input to this machine component).
\end{itemize}
According to the testing strategy devised for stream X-machines \cite{IH97,HI98,IH00}, such a test set can always be constructed for any SXM -- and hence, by extension, for any CSXMS -- that satisfies the relevant DFT conditions.

%% file: 03-sec-rev-psystems.tex
\section{P system models}
\label{sec:rev:psystems}

The P system (membrane system) \cite{handbook} is a model of computation based on eukaryotic cell structures in biology, and the mechanisms used within and between cells to enable communication between their various sub-parts. Since its introduction in \cite{Pau98}, the model has diverged into a number of different variants, each modelling a different combination of biologically-inspired computational mechanisms. In this section we describe a basic variant of the model, and provide a simple example to illustrate its use for computational purposes. We then show how a testing strategy for a system comprising a P system \Base and an SXM \Control can be defined, corresponding to the basic heterotic framework discussed in Sect. \ref{sec:intro}.

\subsection{Cell-like P systems}
Eukaryotic cells are characterised by the presence of membranes, which separate different regions of the cell into a hierarchically organised system of distinct nested compartments. At any given time each compartment will contain a mixture of biochemicals, and this mixture changes over time as a result of the coordinated exchange of biochemicals across membrane boundaries. This basic structure is captured by one of the best known and most utilised types of P system, the \emph{cell-like} P system, using non-cooperative evolution rules and communication rules \cite{GID10}. In the sequel we call these models simply \emph{P systems}.

\begin{definition}\label{def:p-s}
A \emph{P system} with $n$ compartments is a tuple 
$$PS_n=(V,\mu,w_1, \dots, w_n, R_1,\dots,R_n),$$
where
\begin{itemize}
\item $V$ is a finite \emph{alphabet}.
\item $\mu$ defines the \emph{membrane structure}, a hierarchical arrangement of $n$ compartments, identified by integers 1 to $n$.
\item for each $i = 1,\dots,n$, $w_i$ represents the \emph{initial multiset} in compartment $i$.
\item for each $i = 1,\dots,n$, $R_i$ represents the \emph{set of rules} utilised in compartment $i$.
\end{itemize}
\end{definition}
 
The rules capture the way that a chemical species in one cell compartment can be used to trigger the production of new chemical species in both that compartment and others. A typical rule has the form $a \rightarrow (a_1,t_1)\dots(a_m,t_m)$, where $a, a_1,\dots,a_m \in V$ and $t_1,\dots,t_m\in\{here\}\cup\{1, \dots, n\}$. When this rule is applied in a compartment to the symbol $a$, it is replaced in that compartment by the collection of symbols $a_i$ for which $t_i=here$ (by convention, symbols of the form $(a_i, here)$ are often written $a_i$, with the destination being understood). Those symbols $a_i$ for which $t_i=k$ are added to the compartment labelled $k$, provided this is either a parent or a child of the current one. The rules are applied in \emph{maximally parallel mode} in each compartment; for example, if a compartment contains two copies of the symbol $a$, then the rule above will be fired twice (simultaneously), once for each occurrence.


A \emph{configuration} of the P system, $PS_n$, is a tuple $c=(u_1,\dots,u_n)$, where $u_i\in V^*$ for each $i = 1,\dots, n$, which represents the instantaneous disposition of chemical species within the cell's compartments. A \emph{computation} from a configuration $c_1$, using maximally parallel mode, leads to a new configuration $c_2$; this process is denoted $c_1 \Longrightarrow c_2$.

We now discuss, following \cite{GI08}, a testing strategy for P systems which is inspired by the testing principles developed for context-free grammars \cite{L01}, called \emph{rule-coverage}. Other methods for testing P systems also exist, for example mutation-based testing \cite{IG09a}; some are inspired by finite state machine testing \cite{GID10}, others by X-machine testing \cite{IG09b}. Approaches combining verification and testing have also been investigated \cite{GILD10, IGL10}.

%% file: 04-sec-example.tex

\subsection{Rule coverage testing in P systems}
\label{sec:example}

We introduce first some new concepts.
\begin{definition}\label{def:a_rc}
A configuration $c=(u_1, \dots, u_n)$ covers a rule
\[
r_i: a_i \rightarrow (a_{i_1}, j_1)\dots (a_{i_h},j_h)v_i(a_{i_g},j_g)\dots (a_{i_f},j_f)
\]
if there is a computation path starting from the initial configuration and resulting in configuration $c$, during which rule $r_i$ is used. Formally,
\[\begin{aligned}
c_0 &=(w_1, \dots, w_n) \Longrightarrow^* (x_1, \dots, x_{j_1}, \dots, x_{j_h}, \dots, x_ia_i, \dots,   x_{j_g}, \dots,  x_{j_f}, \dots, x_n) \\ 
    &\Longrightarrow
(x_1', \dots, x_{j_1}'a_{j_1}, \dots, x_{j_h}'a_{j_h}, \dots, x_i'v_i, \dots,   x_{j_g}'a_{j_g}, \dots,  x_{j_f}'a_{j_j}, \dots, x_n)\Longrightarrow^*c=(u_1,\dots,u_n)
\end{aligned}\]
\end{definition}

\begin{definition}\label{def:rc}
A \emph{test set}, in accordance to the rule coverage principle, is a set $\Tests^{rc} \subseteq (V^*)^n$, such that for each rule $r\in R_i$, $1 \leq i \leq n$, there is $c \in \Tests^{rc}$ which covers $r$.
\end{definition}

The strategy involved here is to find a test set $\Tests^{rc}$ which unavoidably covers every rule used in the P system, so that when we observe one of the configurations in $\Tests^{rc}$ we can safely deduce that the rule must have been fired during the computation. For example, let us consider the P system with 2 compartments, $PS_2=(V, [[\,]_2]_1,s,t,R_1,R_2)$. This has compartment 2 inside compartment 1, and the alphabet $V$ is the set of symbols that appear in the rules of $R_1$ and $R_2$. Compartment 1 initially contains symbol $s$, compartment 2 contains $t$, and the rules associated with each compartment are
\[\begin{aligned}
R_1 &=\{ r_{11}: s\rightarrow abe; ~~
         r_{12}: a\rightarrow d; ~~
         r_{13}: a\rightarrow c(a,2); ~~
         r_{14}: bc\rightarrow cc; ~~
         r_{15}: e\rightarrow f\}, 
      \text{ and } \\
R_2 &=\{ r_{21}: t\rightarrow b; ~~
         r_{22}: ab\rightarrow c\}.
\end{aligned}\]

A test set for $PS_2$ is $\Tests^{rc}=\{(dbe,b),(ccf,c)\}$, as can be seen from the following two computations:
$$c_0=(s,t) \Longrightarrow^{(r_{11},r_{21})} (abe,b) \Longrightarrow^{(r_{12},\Null)} (dbe,b)$$
and
$$c_0=(s,t) \Longrightarrow^{(r_{11},r_{21})} (abe,b) \Longrightarrow^{(\{r_{13},r_{15}\},\Null)} (cbf,ab) \Longrightarrow^{(r_{14},r_{22})} (ccf,c).$$

One can easily observe that $\Tests^{rc}$ is a test set. All of the rules in both $R_1$ and $R_2$ are covered by at least one element of $\Tests^{rc}$, and there is no way to obtain these configurations without firing each of them at least once.

\subsection{Testing a heterotic P system/SXM system}

We turn now to our first example of heterotic testing. We will assume for this example that \Base is the P system representation of a biocomputational process, while \Control is an SXM representation of a classical digital computer. As prescribed in \cite{KSS+11}, we assume that the biosystem generates an output which the computer inspects; the computer then provides the biosystem with new initial configuration, and the process repeats.

The example above shows that \Base can be tested in isolation, and we saw in Sect. \ref{sec:rev:testing} that general test strategies also exist for testing \Control (subject, in both cases, to certain DFT conditions being satisfied). From a testing point of view, this means that \emph{unit testing} can be assumed to have taken place before the components are combined to form the overall system. The question we now address is how to devise an \emph{integration test} strategy for the combined system. 

\begin{figure}
\centering
\includegraphics*[width=.75\linewidth]{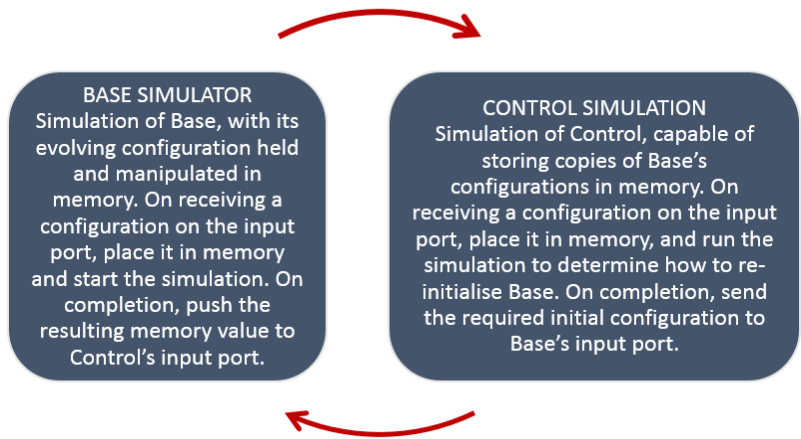}
\caption{A CSXMS that models the interactions between \Base and \Control components in a heterotic system.}
\label{fig:psystem-sxm}
\end{figure}

The simplest approach is to show that \Base and \Control can be represented as components of a CSXMS which models their full combined behaviour. Since this CSXMS is testable, it will follow that the \Base + \Control heterosystem is also testable. Recall that for integration testing purposes, our goal is to test the system generated by composing the P system component (\Base) with the controller (\Control). However, we can easily build communicating SXMs to stepwise-simulate these two agents (Fig. \ref{fig:psystem-sxm}). The \Base simulation holds and manipulates the P system's configurations in memory using an ordinary function that simulates rule execution. Once the computation has run to completion, a second function moves the current memory value (i.e. the final configuration) to the output port, and a communicating function then sends the configuration to \Control. This uses an ordinary function to examine the input port, decides how \Base should be re-initialised, and sends the relevant configuration to its output port. A communicating function then transfers this back to \Base, which uses it as its new initial configuration and the whole process repeats.

Since the simulation of \Base is now part of the CSXMS construction, and we require this to satisfy the stream X-machine DFT conditions, the same should also be true of the P system, and hence of any biological system it may describe. While this may potentially be experimentally unreasonable, we should note that the simulation is doing more work than is required, since we do not need to check, for example, that the P system has computed its terminal configuration correctly (this has already been addressed at the unit testing stage). For integration testing purposes, we can instead regard \Base and \Control as `black boxes', and focus simply on their mutual interactions. In general, abstracting away the components' detailed internal behaviours in this way will considerably simplify the task of ensuring the DFT conditions are satisfied.

%% file: 05-sec-theory.tex
\section{Shortcomings and ongoing research}
\label{sec:shortcomings}

The construction outlined in Fig. \ref{fig:psystem-sxm} is entirely general, provided both \Base and \Control can be stepwise-simulated as components of a CSXMS. This is generally possible, because the underlying SXM model is Turing-complete. However, it is not enough that the components' behaviours can be simulated; it is also important that the simulations are efficient; it would rarely be reasonable, for example, to require companies to build SXM simulations of quantum components in order to test their behaviours as part of a larger system. Apart from the intractability problems that would likely arise, this would introduce a new layer of processing (construction of the simulation itself), which would itself require verification.

As we have noted above, however, the simulations are doing more work than is actually required for integration testing purposes. Indeed, the use of simulations in Fig. \ref{fig:psystem-sxm} above was only introduced for theoretical reasons, to allow us to establish that testability is indeed possible. For practical purposes it would be more sensible to use physical implementations of \Base and \Control as experimental oracles. Instead of simulating a P system, for example, we could instruct \Control to pass details of the next initial configuration to an automated biochemical assembly, causing it to run a physical instantiation of the P system.  Having run the experiment, automated machinery could be used to determine the concentrations of relevant chemicals in the resulting mix, and use these to determine the next signal to be transmitted to \Control's input port (in terms of the formal model, we would modify the example above so that configurations are passed to \Control as elements of \Inputs rather than via the communications port). This approach has the obvious advantage that each component can be implemented in the form in which it was originally manufactured for unit testing, so we can be confident both that the integration and unit test methodologies are consistent with one another, and also that no additional testing is required due to the introduction of an additional simulation stage.

Nonetheless there are situations in which the CSXMS approach proposed above cannot easily be applied in its current form, even ignoring intractability problems that are likely to arise in systems which combine simulable systems to generate non-simulable ones. Following \cite{KSS+11,SAB+12} we have so far assumed the simplest possible design of heterotic system, in which a single \Control unit repeatedly coordinates the configuration of a single \Base unit, and where each unit's computation is allowed to run to completion before control passes to the other. In such a system it is always possible to say which component is running `now', and which will be running `next'. But one can easily envisage situations in which the concept of a `next' computation step is essentially meaningless. For example, consider a future nano-bot system designed to deliver medication to a specific site in a patient's body. One can envisage a scenario in which the bots (\Base) form a swarm of independent magnetically detectable agents, which continually adjust their motion by interacting with the ambient electromagnetic field in their vicinity. To make the system work, an external apparatus monitors the bots' positions in real time, and uses this information to make continuous adjustments to the electromagnetic field surrounding the patient. In such a system the continuity of interaction is an intrinsic part of the specification, and it would not be appropriate to simplify the system by assuming alternate executions of \Base and \Control. Doing so might well allow us to generate a CSXMS-based test set, but it would not allow us to test the intricacies of the system's underlying real-time functionality.

In situations like this, where the concept of a `next computation step' has no meaning, it is not possible to model system changes using the kind of next-state relation associated with automata or stream X-machines. Instead, we need a model in which mutually interacting processes can be defined and combined, no matter whether their operation assumes discrete time, analogue time, linear time, branching time, or even some combination of temporal structures. Our research in this direction is ongoing, and involves the construction of a generalised X-machine model which preserves the essential features of the SXM model, while allowing computations to be defined over arbitrary temporal structures.

Since the relevance of applying a processing function \Proc in a state \State is determined solely by the configuration change induced once traversal of the associated transition has completed, we can describe the transition by a relation $\Trans : 2 \to (\Cfgs \longleftrightarrow \Cfgs)$, where \Cfgs is the set of possible configurations for the SXM in question, and
\begin{equation}
\label{eqn:conditions}
\Trans(0) = \mathrm{id}_{\Cfgs}  \quad\quad \Trans(1) = \under\Proc
\end{equation}
where $\mathrm{id}_{\Cfgs}$ is the identity relation on \Cfgs (we assume $\mathrm{id}_{\Cfgs} \in \Procs$), and $\under\Proc(c) = \{ c' \where c \mathrel{~\vdash_{\Proc}~} c') \}$.

Writing the transition in this way highlights the role of the timing structure, in this case $2 = \{0,1\}$, in determining the effect of firing the transition. Firing a transition changes the configuration from $c \in \Trans(0)(c)$ to some $c' \in \Trans(1)(c)$. If we wish to include instead a continuously evolving analogue procedure for computing \Proc, we can do so formally by replacing the existing transition with any continuous function $\Trans' : [0,1] \to (\Cfgs \longleftrightarrow \Cfgs)$ that also satisfies (\ref{eqn:conditions}). Similarly, transfinite models of computation can be instantiated using $\Times = \beta+1$ for suitable limit ordinals $\beta$ (where we regard $\beta$, the maximal value in $\beta+1$, as the value ``1'' in (\ref{eqn:conditions})).

More generally, given any timing structure, \Times, we can replace any transition in an SXM with a function of the form $\Trans'' : \Times \to (\Cfgs \longleftrightarrow \Cfgs)$ without changing its overall behaviour, provided $\Trans''$ has a minimum element $0$ and maximum element $1$, and satisfies (\ref{eqn:conditions}). Since we are considering physically realisable computations, we also impose the condition that $\Trans''$ should be continuous when regarded as a function on \Times (we regard this as a defining property of what it means for \Times to be a sensible model of time for the computation in question, rather than a constraint on $\Trans''$).

Formally, however, the notion that $\Trans''$ is continuous presupposes the existence of topologies on both \Times and \Cfgs. For philosophical reasons we assume that \Times is partially ordered, and assign it the associated compact Hausdorff topology. Similarly, we can define a natural Tychonov topology on \Cfgs \cite{Sta14}. In this way, we postulate, we can instantiate each transition function using which ever paradigm is most appropriate for the function being modelled, thereby allowing truly general heterotic systems to be brought under the SXM testing umbrella \cite{Sta13,Sta14}. 

%% file: 06-sec-conclusions.tex
\section{Summary and conclusions}
\label{sec:conclusion}

In this paper we have considered the problem of constructing test-sets for integration-testing a heterotic system \XM{H}, composed of two interacting systems, \Base and \Control. For Turing-simulable systems, this can be achieved by re-expressing \Base and \Control as communicating components within a CSXMS model. Since all such models have an associated test-set generation strategy, this allows us to generate adequate test sets for \XM{H}, provided the relevant design-for-test conditions are satisfied. We illustrated our approach by describing how a test set can be generated for a heterotic system combining an automaton-based \Control system with a bio-related P system (\Base). It remains important that these components can also be tested in isolation, and we have seen how unit testing of a P system can be achieved.

It will also be important to validate our method experimentally, since many of the systems we envisage being included in practical heterotic systems cannot be simulated efficiently using traditional SXM-based models, and would be better included as experimental oracles. Such experiments could be conducted both \emph{in silico} and in the laboratory. For example, we can perform various mutation tests on the combined system \XM{H}, by deliberately seeding \Base and \Control with faults and testing our method's ability to detect them.

The technical structure we presented to deduce the existence of a test set is quite general, but while it can easily be generalised to include several interacting components, it cannot cope with situations involving processes whose interactions are sufficiently complicated that the question \emph{what communication event comes next?} is essentially meaningless. In such cases it is necessary to devise an extended model of X-machine computation which is sufficiently general to allow computations and communications with any temporal structure. In this context it is also important to remember that physical systems are invariably noisy, and it will be especially important when devising a fully general testing strategy to ensure that tolerances and thresholds can be specified, and more importantly, tested for. Work on this topic is continuing, and we hope to report positive results in due course.

%% file: 2014-mgmps-ARXIV.bbl
\newcommand{\etalchar}[1]{$^{#1}$}
\begin{thebibliography}{BGG{\etalchar{+}}99}

\bibitem[AB09]{AB09}
J.~Anders and D.~Browne.
\newblock Computational power of correlations.
\newblock {\em Phys. Rev. Lett.}, 102:050502, 2009.

\bibitem[BC96]{BC96}
O.~Bournez and M.~Cosnard.
\newblock On the computational power of dynamical systems and hybrid systems.
\newblock {\em Theoretical Computer Science}, 168(2):417--459, 1996.

\bibitem[BCT12]{BCT12}
E.~J. Beggs, J.~F. Costa, and J.~V. Tucker.
\newblock The impact of models of a physical oracle on computational power.
\newblock {\em Math. Struct. in Comp. Science}, 22:853--879, 2012.

\bibitem[BGG{\etalchar{+}}99]{BCGGHV99}
T.~B\u{a}l\u{a}nescu, H.~Georgescu, M.~Gheorghe, M.~Holcombe, and C.~Vertan.
\newblock Communicating stream {X}-machines are no more than {X}-machines.
\newblock {\em Journal of Universal Computer Science}, 5(9):492--507, 1999.

\bibitem[Cho78]{Cho78}
T.~S. Chow.
\newblock Testing software design modelled by finite state machines.
\newblock {\em IEEE Transactions on Software Engineering}, 4(3):178--187, 1978.

\bibitem[Eil74]{Eil74}
S.~Eilenberg.
\newblock {\em Automata, {L}anguages and {M}achines}, volume~A.
\newblock Academic Press, London, 1974.

\bibitem[GI08]{GI08}
M.~Gheorghe and F.~Ipate.
\newblock On testing {P} systems.
\newblock In D.~W. Corne, P.~Frisco, G.~P{\u a}un, G.~Rozenberg, and
  A.~Salomaa, editors, {\em Membrane Computing}, volume 5391 of {\em Lecture
  Notes in Computer Science}, pages 204--216. Springer Berlin Heidelberg, 2008.

\bibitem[GID10]{GID10}
M.~Gheorghe, F.~Ipate, and C.~Dragomir.
\newblock Formal verification and testing based on {P} systems.
\newblock In G.~P{\u{a}}un, M.~J. P{\'e}rez-Jim{\'e}nez,
  A.~Riscos-N{\'u}{\~n}ez, G.~Rozenberg, and A.~Salomaa, editors, {\em Membrane
  Computing}, volume 5957 of {\em Lecture Notes in Computer Science}, pages
  54--65. Springer Berlin Heidelberg, 2010.

\bibitem[GILD10]{GILD10}
M.~Gheorghe, F.~Ipate, R.~Lefticaru, and C.~Dragomir.
\newblock An integrated approach to {P} systems formal verification.
\newblock In {\em Proceedings of the 11th International Conference on Membrane
  Computing}, CMC'10, pages 226--239, Berlin Heidelberg, 2010. Springer.

\bibitem[HI98]{HI98}
M.~Holcombe and F.~Ipate.
\newblock {\em {Correct Systems: Building a Business Process Solution}}.
\newblock Springer Verlag, 1998.

\bibitem[IBE03]{IBE03}
F.~Ipate, T.~B\u{a}l\u{a}nescu, and G.~Eleftherakis.
\newblock Testing communicating stream {X}-machines.
\newblock In {\em Proceedings of the 1st Balkan Conference in Informatics},
  pages 161--174, 2003.

\bibitem[IG09a]{IG09a}
F.~Ipate and M.~Gheorghe.
\newblock Mutation based testing of {P} systems.
\newblock {\em International Journal of Computers Communications \& Control},
  4(3):253--262, 2009.

\bibitem[IG09b]{IG09b}
F.~Ipate and M.~Gheorghe.
\newblock Testing non-deterministic stream {X}-machine models and {P} systems.
\newblock {\em Electronic Notes in Theoretical Computer Science}, 227:113--126,
  2009.

\bibitem[IGL10]{IGL10}
F.~Ipate, M.~Gheorghe, and R.~Lefticaru.
\newblock Test generation from {P} systems using model checking.
\newblock {\em The Journal of Logic and Algebraic Programming}, 79(6):350--362,
  2010.

\bibitem[IH97]{IH97}
F.~Ipate and M.~Holcombe.
\newblock An integration testing method that is proved to find all faults.
\newblock {\em International Journal of Computer Mathematics}, 63:159--178,
  1997.

\bibitem[IH00]{IH00}
F.~Ipate and M.~Holcombe.
\newblock Generating test sets from non-deterministic stream {X}-machines.
\newblock {\em Formal Aspects of Computing}, 12:443--458, 2000.

\bibitem[IH02]{IH02}
F.~Ipate and M.~Holcombe.
\newblock Testing conditions for communicating stream {X}-machine systems.
\newblock {\em Formal Aspects of Computing}, 13:431--446, 2002.

\bibitem[KSS{\etalchar{+}}11]{KSS+11}
V.~Kendon, A.~Sebald, S.~Stepney, M.~Bechmann, P.~Hines, and R.~C. Wagner.
\newblock Heterotic computing.
\newblock In {\em Unconventional Computation}, volume 6714 of {\em Lecture
  Notes in Computer Science}, pages 113--124. Springer, Berlin Heidelberg,
  2011.

\bibitem[L{\"a}m01]{L01}
R.~L{\"a}mmel.
\newblock Grammar testing.
\newblock In {\em Proceedings of the FASE 2011}, volume 2019 of {\em Lecture
  Notes in Computer Science}, pages 201--216. Springer, Berlin Heidelberg,
  2001.

\bibitem[Lay93]{Lay93}
G.~Laycock.
\newblock {\em {The Theory and Practice of Specification Based Software
  Testing}}.
\newblock PhD thesis, Department of Computer Science, University of Sheffield,
  UK, 1993.

\bibitem[LY96]{LY96}
D.~Lee and M.~Yannakakis.
\newblock {Principles and Methods of Testing Finite State Machines - A Survey}.
\newblock {\em Proceedings of the IEEE}, 84:1090--1123, 1996.

\bibitem[P{\u{a}}u98]{Pau98}
G.~P{\u{a}}un.
\newblock Computing with membranes.
\newblock TUCS Report 208, Turku Centre for Computer Science, 1998.

\bibitem[PRS09]{handbook}
G.~P{\u{a}}un, G.~Rozenberg, and A.~Salomaa, editors.
\newblock {\em {The Oxford Handbook of Membrane Computing}}.
\newblock Oxford Handbooks in Mathematics. OUP, Oxford, 2009.

\bibitem[SAB{\etalchar{+}}12]{SAB+12}
S.~Stepney, S.~Abramsky, M.~Bechmann, J.~Gorecki, V.~Kendon, T.~J. Naughton,
  M.~J. P\'{e}rez-Jim\'{e}nez, F.~J. Romero-Campero, and A.~Sebald.
\newblock Heterotic computing examples with optics, bacteria, and chemicals.
\newblock In J.~Durand-Lose and N.~Jonoska, editors, {\em Unconventional
  Computation and Natural Computation}, volume 7445 of {\em Lecture Notes in
  Computer Science}, pages 198--209. Springer, Berlin Heidelberg, 2012.

\bibitem[Sie99]{Sie99}
H.~T. Siegelmann.
\newblock {\em {Neural Networks and Analog Computation: Beyond the Turing
  Limit}}.
\newblock Birkh\"auser, 1999.

\bibitem[Sta01]{Sta01}
M.~Stannett.
\newblock Computation over arbitrary models of time.
\newblock Tech. Rep. CS-01-08, Dept of Computer Science, University of
  Sheffield, Sheffield, UK, 2001.

\bibitem[Sta13]{Sta13}
M.~Stannett.
\newblock Specification, testing and verification of heterotic computers using
  generalised {X}-machines.
\newblock Poster presentation, Royal Society Workshop: ``Heterotic computing:
  exploiting hybrid computational devices'', Chicheley Hall, 7--8 November
  2013.

\bibitem[Sta14]{Sta14}
M.~Stannett.
\newblock Specification, testing and verification of unconventional
  computations using generalized {X}-machines.
\newblock {\em International Journal of General Systems}, 43(7):713--721, 2014.

\bibitem[TB07]{TB07}
J.~Tucker and E.~Beggs.
\newblock Experimental computation of real numbers by {N}ewtonian machines.
\newblock {\em Proc. R. Soc. A}, 463(2082):1541--1561, 2007.

\end{thebibliography}
